\documentclass[10pt,journal]{IEEEtran}
\usepackage{amssymb}
\usepackage{amsmath}
\usepackage{cite}
\usepackage{url}
\usepackage{bbm}
\usepackage{xcolor}
\usepackage{hyperref}
\usepackage{cite,graphicx,amsmath,amssymb}
\usepackage{fancyhdr}
\usepackage{mdwmath}
\usepackage{mdwtab}
\usepackage{caption}
\usepackage{amsthm}
\usepackage{setspace}
\usepackage{xcolor}
\usepackage{subfig}
\usepackage{multirow}
\usepackage{makecell}
\usepackage{mathtools}
\usepackage[linesnumbered,ruled,vlined]{algorithm2e}

\graphicspath{./src/img/}

\newtheorem{theorem}{Theorem}

\newtheorem{lemma}{Lemma}

\newtheorem{corollary}{Corollary}

\allowdisplaybreaks
\title{Near-Field Sensing: A Low-Complexity Wavenumber-Domain Method}

\author{Hao~Jiang,~\IEEEmembership{Graduate Student Member,~IEEE}, Zhaolin Wang,~\IEEEmembership{Graduate Student Member,~IEEE}, \\ and Yuanwei Liu,~\IEEEmembership{Fellow,~IEEE}

\thanks{ Hao Jiang and Zhaolin Wang are with the School of Electronic Engineering and Computer Science, Queen Mary University of London, London E1 4NS, U.K. (e-mail: \{hao.jiang; zhaolin.wang\}@qmul.ac.uk).

Yuanwei Liu is with the Department of Electrical and Electronic Engineering, The University of Hong Kong, Hong Kong (e-mail: yuanwei@hku.hk).
} 
}

\begin{document}
\maketitle
\begin{abstract}
A novel low-complexity wavenumber-domain method is proposed for near-field sensing (NISE).
Specifically, the power-concentrated region of the wavenumber-domain channels is related to the target position in a non-linear manner. 
Based on this observation, a bi-directional convolutional neural network (BiCNN)-based approach is proposed to capture such a relationship, thereby facilitating low-complexity target localization.
This method enables direct and gridless target localization using only a limited bandwidth and a single antenna array.
Simulation results demonstrate that: 1)~during the offline training phase, the proposed BiCNN method can learn to localize the target with fewer trainable parameters compared to the naive neural network architectures; and 2)~during the online implementation phase, the BiCNN method can spend 100x less time while maintaining comparable performance to the conventional two-dimensional multiple signal classification (MUSIC) algorithms.
\end{abstract}

\section{Introduction}
With the adoption of high-frequency bands and extremely large aperture arrays (ELAAs) in the sixth-generation communication technology (6G), near-field sensing (NISE) has become a hotshot of research \cite{zheng2019localization, yang2024beamfocusing}.
Compared to the conventional angle-only far-field sensing, NISE can facilitate a full capture of the target position, i.e., both angle and distance \cite{wang2023near, liu2023tutorial, he2021mixed}.
Specifically, the curvature of spherical wavefronts in the near-field region causes targets in the same angular direction to experience phase variances.
This feature gives rise to an additional distance dimension hidden in near-field channels.
In fact, by leveraging such a distance dimension, NISE can localize targets with limited bandwidth and a single antenna array \cite{luo20236dradarsensingtracking}.

Given these advantages, significant efforts have been dedicated to harnessing the potential of NISE.
For example, the classic multiple signal classification (MUSIC) algorithm was first utilized for the near-field scenario by \cite{huang1991near}.
The core idea of the extended version of MUSIC was to sample both angle and distance to form two-dimensional (2D) grids. 
Then, a 2D MUSIC spectrum can be constructed by traversing the girds, where the target is estimated to be located at the grid with the highest peak.
Based on the 2D MUSIC, the authors of \cite{li2024near, qu2024two} proposed an angle-then-distance estimation method thereby converting a 2D search into separated one-dimensional (1D) searches.
As an alternative method, the authors of \cite{zhang2018localization} turned 2D search into one 1D search by splitting the directional matrix.
From the viewpoint of electromagnetic theory, authors of \cite{cao2024low} exploited the electromagnetic propagation model to let the receiver find its location with low complexity.
In fact, for most estimation algorithms in near-field sensing, the grid search is necessary due to the highly nonconvex and multimodal property of the resultant optimization problem \cite{pesavento2023marius, ramezani2024localizations}. 
However, the grid search method can lead to high computational complexity, thus resulting in intolerable delays.
To overcome this issue, low-complexity gridless methods are urgently needed for real-world implementations.

Motivated by the above, we propose a low-complexity gridless target localization algorithm for near-field systems.
In particular, due to the positional information in space-domain channel representations being unapparent, we first resort to wavenumber-domain analysis, by which the positional information is related to the power-concentrated region of the channel.
To capture a highly non-linear relationship between the target position and the power-concentrated region, we propose a bi-directional convolutional neural network (BiCNN) design, which enables the estimation of the target's location.
Finally, the simulation results are presented to demonstrate the effectiveness of the proposed method, and further illustrate the superiority of the BiCNN approach over the traditional 2D-MUSIC algorithm in terms of complexity.

\section{System Model}\label{sect:system_model}
\begin{figure}
    \centering
    \includegraphics[width=0.8\linewidth]{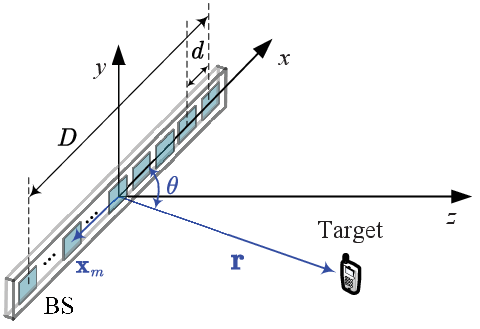}
    \caption{An illustration of the system model.}
    \label{fig:syst_model}
\end{figure}
As illustrated in Fig. \ref{fig:syst_model}, we consider a near-field multiple-input single-output (MISO) system, which encompasses a base station (BS) with a $M$-antenna uniform linear array (ULA), where $M=2\tilde{M}+1$ and a single-antenna target.
Here, the full-duplex technique is utilized at the BS, allowing for simultaneous transmission and reception.
The transceivers are assumed to be located on the $xz$ plane, and the ULA at the BS is placed along the $x$-axis. 
Letting the carrier frequency and the wavelength be $f$ and $\lambda$, respectively, the array aperture of the ULA at the BS can be calculated by $D=(M-1)d$, where $d=0.5\lambda$.
Here, letting $r$ and $\theta$ be the Euclidean distance between the BS and the target and the angle between the target and the $x$-axis respectively, it is assumed that the target is in the radiating near-field region of the BS, which is defined by $ r < \frac{2D^2}{\lambda}$.
Please note that the reactive near-field region is omitted here, since it is confined within a few wavelengths \cite{ouyang2024impact}.
One narrow band with width $B$ is utilized for localization of the target, and the channel reciprocity is assumed to hold in our system.

Due to the inaccuracy of planar-wave assumption in near-field region\cite{liu2023tutorial}, we adopt a dedicated near-field channel to capture the characteristic of the channel.
Without loss of generality, we set the origin of the coordinate system at the center of the ULA.
Hence, the coordinate of the $m$-th antenna on the ULA and that of the target can be expressed as $\mathbf{x}_{m} = [x_m^{(x)}, x_m^{(y)}, x_m^{(z)}] \in \mathbb{R}^{3 \times 1}$ and $\mathbf{r} = [r^{(x)}, r^{(y)}, r^{(z)}] \in \mathbb{R}^{3 \times 1}$, respectively.
Subsequently, the near-field array response for a given position $\mathbf{r}$ can be expressed as 
\begin{align}
    \mathbf{a}(\mathbf{r})=\left[ e^{-jk_0 \left\| \mathbf{r} - \mathbf{x}_{-\tilde{M}} \right\|_2},...,e^{-jk_0 \left\| \mathbf{r} - \mathbf{x}_{+\tilde{M}} \right\|_2} \right] ^{\mathrm{T}}, \label{eq:array_resp_single}
\end{align}
where $\left\| \cdot \right\| _2$ denotes the two-norm of a vector, and $k_0 \triangleq \frac{2 \pi}{\lambda}$ denotes the wavenumber.
Therefore, according to the channel reciprocity, a round-trip channel matrix for position $\mathbf{r}$ can be expressed by
\begin{align}
    \mathbf{H}= \beta\mathbf{a}(\mathbf{r}) \mathbf{a}^{\rm T}(\mathbf{r}). \label{eq:array_resp_round}
\end{align}
where $\beta $ denotes the channel gain for the line-of-sight (LoS) channel.
Here, the LoS channel gain can be calculated by $\beta \triangleq \mathrm{\zeta}_{\mathrm{pathloss}}\left( f, 2r \right) G_{\mathrm{t}}G_{\mathrm{r}}$, where $G_{\mathrm{t}}$ and $G_{\mathrm{r}}$ denote the transmit and receive array gain, respectively, and $\mathrm{\zeta}_{\mathrm{pathloss}}^{}\left( f,d \right) =\sqrt{\frac{c}{4\pi f}}\frac{1}{d}$ denotes the pathloss with $c$ denotes the speed of light. 
It is noted that, since the target is assumed to be within the radiating near-field region, the pathloss variances among antenna elements can be ignored and replaced by the pathloss of the center link \cite{wang2023near}.

Based on the modeling above, letting $s[n] \in \mathbb{C}$ be the probing signal at the BS during the $n$-th coherent time block, the transmitted signal can be expressed as 
\begin{align}
    \mathbf{x}[n] = \sqrt{P}\mathbf{w}s[n],
\end{align}
where $P$ denotes the transmit power at the BS, and $\mathbf{w} \in \mathbb{C}^{M \times 1}$ is the precoding vector, with unit power constraint $\left| \mathbf{w}^{\mathrm{H}}\mathbf{w} \right|=1$.
After propagating through the round-trip channel, the received echo signal at the BS can be expressed as 
\begin{align}
    \mathbf{y}[n] = \mathbf{H}\mathbf{x}[n] + \mathbf{z}[n],
\end{align}
where $\mathbf{z}[n] \in \mathcal{CN}(\mathbf{0}_M, \sigma^2\mathbf{I})$ denotes the complex Gaussian noise with zero mean and power $\sigma^2$, and $\mathbf{I}$ denotes a identity matrix.

At high-frequency bands, the wireless channels are LoS-dominated and non-line-of-sight (NLoS) assisted, as a result of severe scattering loss \cite{tang2024line}.
Therefore, in this work, we deal with the LoS case.
Since the position of the target is embedded in the LoS channel, target localization is possible.
However, the position is highly spread and unapparent in the space-domain channel $\mathbf{H}$, thus entailing complex signal processing algorithms to extract the position. 
To circumvent the high computational complexity introduced by the adoption of these algorithms, we transform the space-domain channel representation into the wavenumber domain, where the position can be apparently observed from the wavenumber domain.

\section{Wavenumber-Domain Localization Algorithm}
In this section, we first introduce how to transform the space-domain channel into the wavenumber-domain, where the power-concentrated region of the channel is related to the target's position.
Then, we elaborate a learning algorithm to learn such a relation, thus facilitating localization.

\subsection{Wavenumber-Domain Transform}
The core idea of wavenumber-domain transformation is to approximate the spherical wavefronts in the near-field region with a summation of planar wavefronts, whose form is coincidental with the four-dimensional (4D) Fourier transformation.
Deviated from the canonical wavenumber-domain transformation presented by  \cite{pizzo2022fourier}, the transceivers are co-located at the same position for our case, due to the full-duplex antenna architecture. 
Therefore, by converting the 4D Fourier transformation into a 2D counterpart, the entry at the intersection of the $i$-th column and $j$-th row in $\mathbf{H}$ can be written as
\begin{align}
    \left[ \mathbf{H} \right] _{i,j}=\frac{1}{4\pi^2}\iint_{\mathcal{D} _{\mathbf{k}}}{a\left( \mathbf{k},\mathbf{x}_m \right) h_a\left( k_x,k_y \right) a^*\left( \mathbf{k},\mathbf{x}_m \right) dxdy}, \label{eq:wave_number_domain_integral}
\end{align}
where $h_a\left( k_x, k_y \right)$ denoted the coupling coefficient between transmit/receive array response $a\left( \mathbf{k},\mathbf{x}_m \right)$, and integral region $\mathcal{D}_{\mathbf{ k}}$ can be specified by $\mathcal{D} _{\mathbf{k}}\triangleq \left\{ \left( k_x,k_y \right) \in \mathbb{R} ^2:~k_x^2+k_y^2\le k_{0}^{2} \right\}$.
In addition, the array response are specified by $a\left( \mathbf{k},\mathbf{x}_m \right) = \exp \left\{ -j\mathbf{k}^{\mathrm{T}}\mathbf{x}_m \right\}$, where $\mathbf{k} \triangleq [k_x, k_y, \gamma(k_x, k_y)]^{\rm T}$ and $\gamma(a, b) \triangleq \sqrt{k_0^2 - a^2- b^2}$.
To make \eqref{eq:wave_number_domain_integral} more tractable, we discretize the integral region $[-k_0, +k_0]$ by sampling it with an interval $2\pi/D$.
Moreover, since the whole system is located on the $xz$ plane, the $y$-coordinate can be discarded.
Jointly considering the two factors, the discretized region can be expressed as 
\begin{align}
    \mathcal{G} _{\mathbf{k}}\triangleq \left\{ \varepsilon \in \mathbb{Z} :-\lceil {Dk_0}/{2\pi} \rceil \le \varepsilon \le +\lfloor {Dk_0}/{2\pi} \rfloor \right\},
\end{align}
where $\lceil \cdot \rceil$ and $\lfloor \cdot \rfloor$ denote the ceiling and flooring functions, respectively.
Accordingly, by denoting the cardinality of $\mathcal{G} _{\mathbf{k}}$ as $|\mathcal{G} _{\mathbf{k}}|$, the integral form in \eqref{eq:wave_number_domain_integral} can be converted to a summation form (equivalently matrix multiplication):
\begin{align}
    \mathbf{H}\overset{\rm (a)}{\simeq} M\sum_{i=1}^{|\mathcal{G} _{\mathbf{k}}|}{\sum_{j=1}^{|\mathcal{G} _{\mathbf{k}}|}{\mathbf{a}\left( i \right) \left[ \mathbf{H}_a \right] _{i,j}\mathbf{a}^*\left( j \right)}}=M\mathbf{AH}_a\mathbf{A}^{\mathrm{H}},
\end{align}
where $\mathbf{H}_a$ denotes the wavenumber-domain channel representation, and the array response can be written as 
\begin{align}
    \mathbf{a}\left( i \right) =\frac{1}{\sqrt{{M}}}\left[ \exp \left\{ -j\frac{2\pi i}{D}x_{-\tilde{M}}^{\left( x \right)} \right\} ,...,\exp \left\{ -j\frac{2\pi i}{D}x_{+\tilde{M}}^{\left( x \right)} \right\} \right] ^{\mathrm{T}}.
\end{align}
Please note that the approximation of $\rm (a)$ is very accurate for a normalized size $D$ larger than a few wavelength $\lambda$ \cite{pizzo2020spatial}.
Since the wavenumber-domain transformation matrix (WTM) $\mathbf{A} \triangleq [\mathbf{a}\left( 1\right), ..., \mathbf{a}\left( |\mathcal{G} _{\mathbf{k}}| \right)]$ is semi-unitary, indicating that $\mathbf{A}^{\rm H}\mathbf{A}=\mathbf{I}$ holds, we can obtain the wavenumber-domain channel via:
\begin{align}
    \mathbf{H}_a=\frac{1}{M}\mathbf{A}^{\mathrm{H}}\mathbf{HA}.
\end{align}
It is important to note that the wavenumber-domain channel $\mathbf{H}_a$ is obtained by sampling wavenumber $\mathbf{k}$, which is directional.
Therefore, the power-concentrated region of $\mathbf{H}_a$ is correlated with the direction of the signal.
More importantly, in near-field systems, the non-negligible aperture size causes the distance dimension to spread across different directions, which in turn facilitates the acquisition of distance information.
Therefore, we will present a learning algorithm to learn this relation, 

\subsection{Target Localization Algorithm}
\subsubsection{Prepossessing} In the first stage, the BS will transmit a probing signal that is specially designed to obtain the target's location.
In this work, we utilize only a single snapshot (or, equivalently, a coherent time block) for localization.
Therefore, for the brevity of notations, we drop the index of the coherent time block.
It is worth noting that incorporating multiple snapshots can enhance the performance of our method, as averaging over several snapshots helps mitigate the impacts of noise.
We utilize $\mathbf{w}^\prime =\mathbf{A}\mathbf{1}$ to transmit unit power probing signal $s$, where $\mathbf{1}\in\mathbb{R}^{M \times 1}$ is an all-one vector.
Under the unit power constraint, the transmit beamformer can be obtained by $\mathbf{w}=\mathbf{w}^\prime / \left\| \mathbf{w}^\prime \right\| _2$.
After reception at the BS, the echo signal $\mathbf{y}$  is processed by a matrix multiplication by $\mathbf{A}^{\rm H}$, thus yielding 
\begin{figure}[t!]%
    \centering
    \subfloat[Observation vector 
 $\mathbf{o}$ when $r=10~m$]{
        \includegraphics[width=0.7\linewidth]{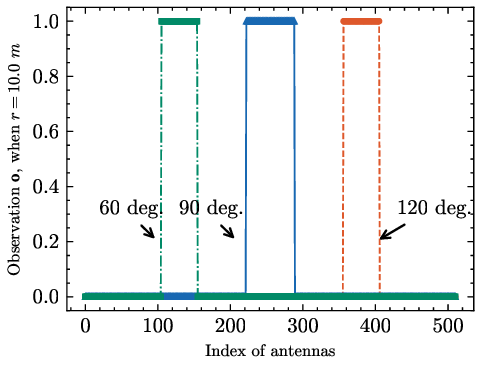}
        } \\
    \subfloat[Observation vector $\mathbf{o}$ when $r=30~m$]{
        \includegraphics[width=0.7\linewidth]{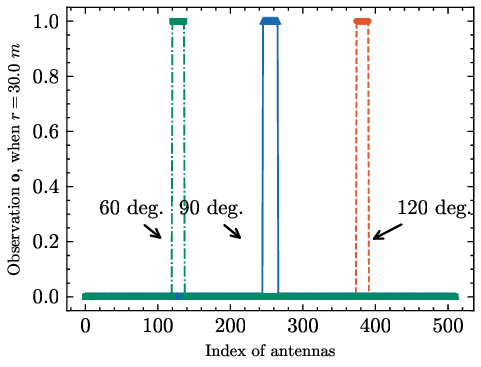}
        }
    \caption{An illustration of the received observation under different distances and angles of the target without noise and pathloss.}
    \label{fig:observation}
\end{figure}
\begin{align}
    \tilde{\mathbf{o}} &=\frac{\mathbf{A}^{\mathrm{H}}\mathbf{y}}{s} =\frac{\sqrt{P}\mathbf{A}^{\mathrm{H}}\mathbf{Hw}^{\prime}}{\left\| \mathbf{w}^{\prime} \right\| _2}+\frac{\mathbf{A}^{\mathrm{H}}\mathbf{z}}{s} \notag\\ &{=}\frac{\sqrt{P}\mathbf{A}^{\mathrm{H}}\mathbf{AH}_a\mathbf{A}^{\mathrm{H}}\mathbf{A}\mathbf{1}}{\left\| \mathbf{w}^{\prime} \right\| _2}+\frac{\mathbf{A}^{\mathrm{H}}\mathbf{z}}{s}\overset{(\rm a)}{=}c\mathbf{H}_a\mathbf{1}+\tilde{\mathbf{z}},
\end{align}
where $\tilde{\mathbf{o}}$ is defined as the observation vector; step $\rm (a)$ is achieved by leveraging the semi-unitary property of WTMs; and $c=\frac{\sqrt{P}}{\left\| \mathbf{w}^{\prime} \right\| _2}$ is a constant; and $\tilde{\mathbf{z}} \triangleq \frac{\mathbf{A}^{\mathrm{H}}\mathbf{z}}{s}$.
To extract power patterns in $\tilde{\mathbf{o}}$ and normalize the observation to range $[0, 1]$, each entry of the normalized observation vector $\mathbf{o}\in \mathbb{R}^{M \times 1}$ can be calculated using
\begin{align}
    \left[ \mathbf{o} \right] _i=\mathbbm{1} \left\{ \frac{\left[ \tilde{\mathbf{o}}^{|\cdot |} \right] _i-\min \left\{ \left[ \tilde{\mathbf{o}}^{|\cdot |} \right] _i \right\}}{\max \left\{ \left[ \tilde{\mathbf{o}}^{|\cdot |} \right] _i \right\} -\min \left\{ \left[ \tilde{\mathbf{o}}^{|\cdot |} \right] _i \right\}}>0.5 \right\}, \label{eq:normalization}
\end{align}
where $i \in \{1, 2, ..., M\}$; $\mathbbm{1}\{\mathrm{condition}\}$ is the indicator function, whose output is 1 when $\mathrm{condition}$ is true and 0 otherwise; and ${(\cdot)}^{|\cdot |}$ denotes the element-wise modulus function.
With normalization in \eqref{eq:normalization}, each of the entries is mapped to either 0 or 1, thus facilitating training by circumventing the bias introduced by pathloss and noise.

Fig. \ref{fig:observation} depicts the received observation vector when the target is located at different positions.
As we can see, the target's angle is related to where the power concentrates in $\mathbf{o}$, while the target's distance is correlated to the width of the power concentrated region.
To capture this non-linear relationship between the power-concentrated region and the location of the target, we resort to machine learning tools.

\subsubsection{Neural Network (NN) Architecture}
\begin{figure}[t!]
    \centering
 \includegraphics[width=0.5\linewidth]{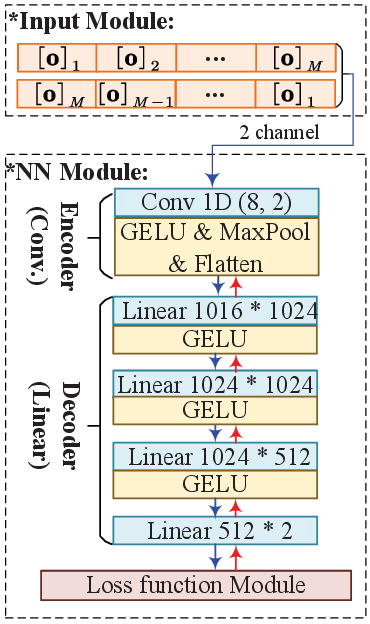}
    \caption{An illustration of the proposed BiCNN method.}
    \label{fig:CNN_archi}
\end{figure}

Since this feature of the observation vector is localized in the power-concentrated region, we, therefore, employ a convolutional neural network (CNN)-based method to capture the feature's locality.
The overall CNN-based algorithm is shown in Fig. \ref{fig:CNN_archi}.
For the input layer, we propose a bi-directional architecture in order to let the NN learn the positional distribution better.
Specifically, the input of the NN can be obtained by 
\begin{align}
    \mathbf{O} = \mathrm{Stack}\{\mathbf{o},  \mathbf{o}^\prime\} \in \mathbb{R}^{2 \times M},
\end{align}
where $\mathrm{Stack}\{\mathbf{a}, \mathbf{b}\}$ denotes the operation that same-dimension vectors $\mathbf{a}$ and $\mathbf{b}$ are stacked vertically, and $\mathbf{o}^\prime$ denotes the flipped version of $\mathbf{o}$.
Then, the input will be fed to the Encoder, which is processed by a one-dimensional convolutional kernel.
The Gaussian error linear units ($\mathrm{GeLU}$) function is utilized for activation, which can avoid dead neuron issue of the conventional ReLU activation \cite{hendrycks2023gaussianerrorlinearunits}. 
In addition, the $\mathrm{MaxPool}$ is utilized for downsampling the feature map, whose output is further flattened as a feature vector.
In the sequel, the feature vector will be processed by the Decoder, which consists of a fully connected structure, and finally mapped to a two-dimensional estimated target's location vector, i.e., $\hat{\mathbf{r}} = [\hat{x}, \hat{y}]^{\rm T}$.
To train the NN, we utilize Huber loss as the loss function for estimation error, which is defined by 
\begin{align}
    \mathcal{L} _{\boldsymbol{\phi }}^{(1)}\left(\mathbf{r},\hat{\mathbf{r}};\rho \right)  =\begin{cases}
	0.5\left\| \mathbf{r} -\hat{\mathbf{r}} \right\| _{2}^{2},~~\mathrm{if}~\left\| \mathbf{r}-\hat{\mathbf{r}} \right\| _{2}^{}\le \rho\\
	\rho \left\| \mathbf{r}-\hat{\mathbf{r}} \right\| _{2}^{}-0.5\rho , ~~\mathrm{otherwise}\\ 
\end{cases}\label{eq:hbloss}
\end{align}
where $\rho$ is a hyper parameter and $\boldsymbol{\phi }$ denotes the vector containing all the trainable parameter.
The core idea of \eqref{eq:hbloss} is: 1)~when the estimation error is small, i.e., $\left\| \mathbf{r}-\hat{\mathbf{r}} \right\| _{2}^{}\le \rho$, a quadratic penalty will be imposed to force the NN to reduce the small errors; and 2) otherwise, a linear penalty will be imposed to make NN less susceptible to large outliers.
By balancing the sensitivity to outliers via $\rho$, the optimization landscape can be smoother compared to the mean squared error (MSE) and the mean absolute error (MAE) criteria.
Further, to circumvent the over-fitting problem, we introduce L2 regularization to the loss function, i.e.,
\begin{align}
    \mathcal{L} _{\boldsymbol{\phi }}^{(2)}=\mu \sum\nolimits_{i\in \mathcal{I}}^{}{\left[ \boldsymbol{\phi } \right]}_i, \label{eq:l1loss}
\end{align}
where $\mu>0$ denotes the weight of this term.
In fact, $\mathcal{L} _{\boldsymbol{\phi }}^{(2)}$ in \eqref{eq:l1loss} will penalize more complexity of the NN, thus overcoming the overfitting issue.
Jointly considering \eqref{eq:hbloss} and \eqref{eq:l1loss}, the overall loss function and the corresponding parameter updating rule can be written as 
\begin{align}
    \mathcal{L} _{\boldsymbol{\phi }}^{}&=\mathcal{L} _{\boldsymbol{\phi }}^{(1)}\left( \mathbf{r}, \hat{\mathbf{r}};\rho \right) +\mathcal{L} _{\boldsymbol{\phi }}^{(2)},\\
    \boldsymbol{\phi }&=\boldsymbol{\phi }-w\nabla \mathcal{L} _{\boldsymbol{\phi }}^{},
\end{align}
where $w>0$ denotes the learning rate.
Therefore, the whole BiCNN can be seen as a function that maps the observations to the target location, i.e., $\mathcal{F}(\cdot; \boldsymbol{\phi}) : \mathbb{R}^{2 \times M} \mapsto \mathbb{R}^{2 \times 1}$.

\begin{algorithm}[t!]
    \SetAlgoLined 
	\caption{BiCNN-Based Target Localization}\label{alg:1}
	\KwIn{Wavenumber-domain transformation matrix $\mathbf{A}$ and pre-trained BiCNN $\mathcal{F}(\cdot; \boldsymbol{\phi})$. } 
	\KwOut{Estimated position of the target $\tilde{\mathbf{r}}$.}
	\tcp{Transmit Probing Signal:}
    Transmit symbol $s$ using beamformer $\mathbf{w} =\mathbf{A}\mathbf{1} / \left\|\mathbf{A}\mathbf{1} \right\|_2 $, and receive the echo signal $\mathbf{y}$\;
    \tcp{Echo Signal Prepossessing:}
    Combine the echo signal $\mathbf{y}$ using $\mathbf{A}^{\rm H}$ to obtain the raw observation $\tilde{\mathbf{o}}$\;
    Normalize $\tilde{\mathbf{o}}$ using \eqref{eq:normalization} to obtain ${\mathbf{o}}$\;
    Stack the $\tilde{\mathbf{o}}$ and a flipped version ${\mathbf{o}}^\prime$ to form $\mathbf{O}$\;
    \tcp{BiCNN Estimation Step:}
    Feed $\mathbf{O}$ into the pre-trained BiCNN $\mathcal{F}(\cdot; \boldsymbol{\phi})$ to obtain the estimated target's location $\hat{\mathbf{r}}$\;
    \Return{Estimated target's location $\hat{\mathbf{r}}$}
\end{algorithm}

\subsubsection{Training \& Testing of BiCNN}
In general, we adopt the offline training and online implementation scheme, aiming at lowering the computational complexity.
In the training process, we adopt an exponentially decaying learning rate, indicating the learning rate will decay by $\alpha$ after each epoch.
By doing so, the trainable parameters can be updated in a coarse-to-fine manner.
Therefore, the learning rate transition from the $k$-th to the ($k+1$)-th epoch can be expressed as $w_{k+1}=\alpha w_k$.
After being offline trained, the implementation procedure is summarized by \textbf{Algorithm \ref{alg:1}} 

\section{Simulation Results} \label{sect:simulation_res}
This section presents the numerical results that validate the effectiveness of the proposed BiCNN method. The following parameter settings are consistently applied throughout our simulations unless otherwise specified.
For physical-layer parameters, the transmit power is set to $P=30$ dBm, while the noise power spectral density is $-174$ dBm/Hz.
The system functions at $28$ GHz, with a $10$ kHz bandwidth.
The BS is assumed to be equipped with $M=511$ antennas with half-wavelength spacing, and the number of targets is one.
Array gains are set to $G_{\rm t}=10^{1.5}$ and $G_{\rm r}=10^{0.5}$.
For the training process of BiCNN, its structure is illustrated by Fig. \ref{fig:CNN_archi}, in which $\rm Conv 1D (8, 2)$  means a one-dimensional convolution operation with $8$ output channels with a kernel size $2$, and $\mathrm{Linear}~M * N$ indicates a linear layer with $M$-dimensional input and $N$-dimensional output.
We discrete an angle region $[\pi/4, 3\pi/4)$ with an interval $0.01$ and a distance region $[8~\mathrm{m}, 35~\mathrm{m}]$ with an interval of $0.01$ to generate dataset, in which $70\%$, $20\%$, and $10\%$
are utilized for training, validation, and testing, respectively.
Adaptive moment estimation (Adam) is used for optimization with a learning rate  $w=0.001$ with a decaying factor $\alpha=0.98$.
The rest hyper-parameters are set to $\mu=10^{-5}$ and $\rho=1.0$.

\begin{figure}[t!]
    \centering
    \includegraphics[width=0.8\linewidth]{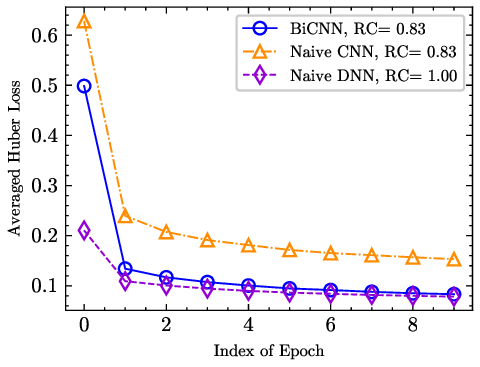}
    \caption{Learning progress of BiCNN compared to the benchmarks}
    \label{fig:train_process}
\end{figure}
In Fig. \ref{fig:train_process}, we compare the training process of the proposed BiCNN method with other benchmarks, i.e., ``Naive CNN" and ``Naive deep neural network (DNN)".
In addition, ``RC" in the legend of Fig. \ref{fig:train_process} represents the number of relative trainable parameters that each NN has compared to the DNN.
This metric is directly related to the training complexity.
Compared to the `Naive CNN", which doesn't include a flipped observation vector as an input channel, BiCNN can capture the locality of the observation and achieve a lower Huber loss (or equivalently higher accuracy).
Compared to the ``Naive DNN", which replaces the convolution encoder with a linear layer, BiCNN can compete with accuracy, while with a lower number of trainable parameters, thereby striking a good balance between performance and complexity.

\begin{figure}[t!]
    \centering
    \includegraphics[width=0.8\linewidth]{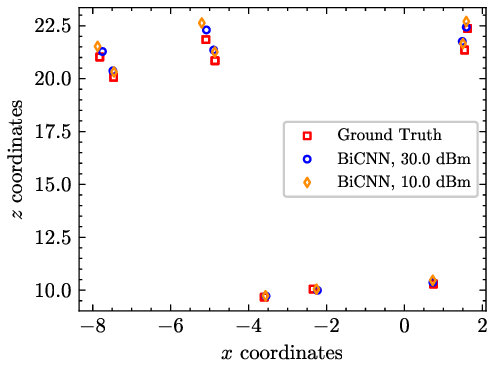}
    \caption{An illustration of BiCNN's performance on the real-channel with different transmit power through 100 Monte Carlo simulation.}
    \label{fig:testing_results}
\end{figure}
In Fig. \ref{fig:testing_results}, the performance of the pre-trained BiCNN on the real channel, where noise and pathloss are included.
This figure depicts the generalization ability of our method.
As we can observe from this figure, the proposed BiCNN can achieve a high estimation accuracy, without traversing any pre-defined grids. 
However, as the distance between BS and the target grows larger, the estimation error will increase.
In addition, as the transmit power increases by $20$ dBm, the performance can be slightly elevated, which shows the good robustness of BiCNN to the variance of transmit power.

In Tab. \ref{tab:1}, we compare the conventional 2D MUSIC algorithm with the proposed BiCNN method.
The MUSIC algorithm follows a two-step procedure: 1)~obtaining the signal and noise subspaces with a computational complexity $\mathcal{O}(M^3)$, and 2)~finding peaks on a discrete 2D coordinate system with a computational complexity dependent on the number of grids.
In this table, RMSE stands for rooted mean square error of the localization, and all the results are obtained through 100 Monte Carlo simulations.
As shown by Tab. \ref{tab:1}, the proposed BiCNN method can eliminate the eigendecomposition in the first step and the grid searching in the second step of MUSIC.
More importantly, our method can achieve decent accuracy while significantly reducing the running time.
\begin{table}[!t]
    \centering
    \caption{Comparison with 2D MUSIC algorithm with $P=30$ dBm.} \label{tab:1}
    \begin{tabular}{cccc}
        \hline
        \textbf{Method} & \textbf{\# Grids} & \textbf{RMSE}  &  \textbf{Avg. Run Time} \\
        \hline
        \multirow{3}{*}{MUSIC} & 10 & 3.8925 m  & 0.4510 s \\
        \cline{2-4} & 100 & 0.6786 m  & 0.4270 s\\
        \cline{2-4} & 1000 & 0.2109 m & 0.5565 s\\
        \hline
        BiCNN & N/A &  0.2439 m & 0.0037 s \\
        \hline
    \end{tabular}
\end{table}

\section{Conclusion} \label{sect:conclusion}
A low-complexity wavenumber-domain method has been proposed for near-field localization.
Specifically, gridless BiCNN is utilized to map the non-linear relationship between the observation and the target's location.
In addition, a bi-directional structure is proposed to boost NN's performance with lower trainable parameters.
Simulation results demonstrate that BiCNN can achieve a decent localization accuracy with much lower computational complexity, compared to conventional 2D MUSIC.
The generalization of this method can be a promising direction for future research.

\bibliographystyle{IEEEtran}
\bibliography{reference/mybib}
\end{document}